# Apoptosis classification using attention based spatio temporal graph convolution neural network

Akash Awasthi

Department of ECE

University of Houston

**Abstract:**

Accurate classification of apoptosis plays an important role in cell biology research. There are many state-of-the-art approaches which use deep CNNs to perform the apoptosis classification but these approaches do not account for the cell interaction. Our paper proposes the Attention Graph spatio-temporal graph convolutional network to classify the cell death based on the target cells in the video. This method considers the interaction of multiple target cells at each time stamp. We model the whole video sequence as a set of graphs and classify the target cell in the video as dead or alive. Our method encounters both spatial and temporal relationships.

## 1 Introduction:

Apoptosis is the programmed cell death event which occurs in multicellular organisms [1]. Apoptosis includes membrane blebbing, cell shrinking, deoxyribonucleic acid (DNA) degradation and in some cases the formation of apoptotic bodies depending on the size of the target cells, [2]. The process of Automatic detection or classification of apoptosis has been in great need recently following the development of high throughput screening assays [3]– [6]. More recently, with the advancements in immunotherapy for the treatment of cancer, single-cell methods like TIMING (time-lapse microscopy in nanowell grids) [7], [8] enable the monitoring of interactions between immune cells [9], [10] and cancer cells in hundreds of thousands of sub-nanoliter wells. These assays have been used to understand the biology of the immune cells and their interaction with tumor cells in a number of preclinical settings [11], [12]. Traditional methods using biochemical assays [13] to tag apoptotic cells are widely used, such as fluorophore conjugated Annexin V [14]–[16] and SYTOX [17]. However, cytotoxicity of chemical assays and phototoxicity due to exposure to light in fluorescent microscopy could adversely affect cell behaviors and lead to cell death [18]. There are some deep learning methods based on CNN but these do not account for cell interactions among the different cells. In this method, we have approached the apoptosis classification problem using the spatio temporal graph, since spatio temporal graphs account for the spatial as well as the temporal relationships. We have used the attention-based graph convolution network to classify the cell death. We have modeled the video frames of the cell interaction as a sequence of the graphs, with each node representing a particular target cell in the video. Below fig (1) represents the video with the three frames having one target cell

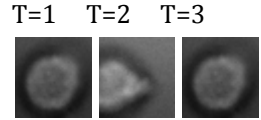

Fig (1) : Target cell for 3 time steps

We have encoded these images of each cell for each time using the trained Resnet50 and assigned these embeddings to each node of the graph. Attention based graph convolution network is used to perform the node classification. Each cell in the video is assigned a label based on the three consecutive death cell events. A label is assigned for the whole sequence of the graph for each node which represents the cell.

## 2 Literature survey:

CNNS are not able to capture the multiple cell interactions between the multiple cells in the videos. Graph neural networks are adopted for any time series classification and regression tasks. In our case our data is a video based which is nothing but the sequence of multiple images with respect to time. Spatio temporal graph neural networks are used for many spatio temporal forecasting tasks (traffic forecasting). Graph neural networks are used to solve many medical or health care related problems like understanding the protein-to-protein interactions.
 GraphGR, a graph neural network, is developed with sophisticated attention propagation mechanisms to predict the therapeutic effects of kinase inhibitors across various tumors[19]. GNNs are used for many pandemic forecasting tasks to predict the number of COVID cases in the future by involving the connectivity between the multiple countries.

## 3 Methodology:

### 3.1 Data preparation:

1) **TIMING Pipeline to extract the cells**

Our dataset is processed using the TIMING pipeline. It is a set of algorithms for nanowell detection, cell segmentation and cell tracking. We have used this pipeline to extract a single cell status during each time stamp. This pipeline classifies each cell into their categories effector or the target cell and gives the Tracking coordinates for each cell. We then use these coordinates to extract the cell from the each timestep image. For this study, we have considered the 15 time points or ( time frames in the video ) for training and testing the graph neural network. Channel 1 is used to extract the Target cells and we are just limited to the target for this study. The TIMING pipeline is executed for the 30 timeframes but later we have divided each video into 2 videos of 15 frames to increase the data set size. This helps in generalization of the model.

2) **Labeling the cells:** After extracting the target cells for each time point from each nanowell image, each target cell during each time point is labeled using the threshold. We have used fluorescent channel ( channel 3 ) to check if the death marker is greater than the threshold then the target cell is assigned as a death cell otherwise alive.

### 3.2 Modeling the dataset as spatio-temporal graphs:
In this apoptosis classification , we have the video of cells interacting with each other. And video is the collection of the multiple frames or images which corresponds to each time stamp. Therefore, video data consist of spatial and temporal relationships. We have modeled the video data as a Spatio-temporal graph where each frame corresponds to a graph and cells in each frame represent the number of nodes ( N ). A graph can be represented as G(E,N ) where N is the number of nodes and the E is the number of edges. Each cell in a particular frame is considered as a node. We have considered the video of size 15, therefore we have 15 graphs representing each frame or time stamp of the video. **Number of cells in each video is different.** For our dataset, we have a maximum of 3 cells in each video. Therefore, we have fixed the number of nodes across each sequence of graphs as 3. Each node represents a cell. If there are less than 3 cells in the video, 0 vector is assigned to that particular node. Initially, we have assumed all the graphs as fully connected with the edge weights unity.

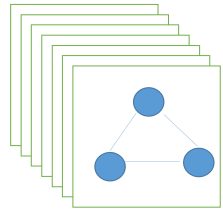

**fig(2) Video as a sequence of the graph**

In this problem, we are performing spatio temporal node based classification. Sequence length of 15 is considered for each spatio temporal graph which represents one of the video samples. We have assigned the label to the cell of the video based on the three consecutive times in the video if the cell has died. Spatio-temporal based Graph Neural networks are important with the time series or video based dataset since these networks capture spatial and temporal relationships. Each sample video consists of one more than one cell interacting with each other. We have a graph representing each time stamp. And time stamps have more than one cell interacting with each other. Resnet architecture[20] is trained on the cell images with their labels and the image embedding is obtained for each image using the trained resnet architecture to assign the node features in the graph.

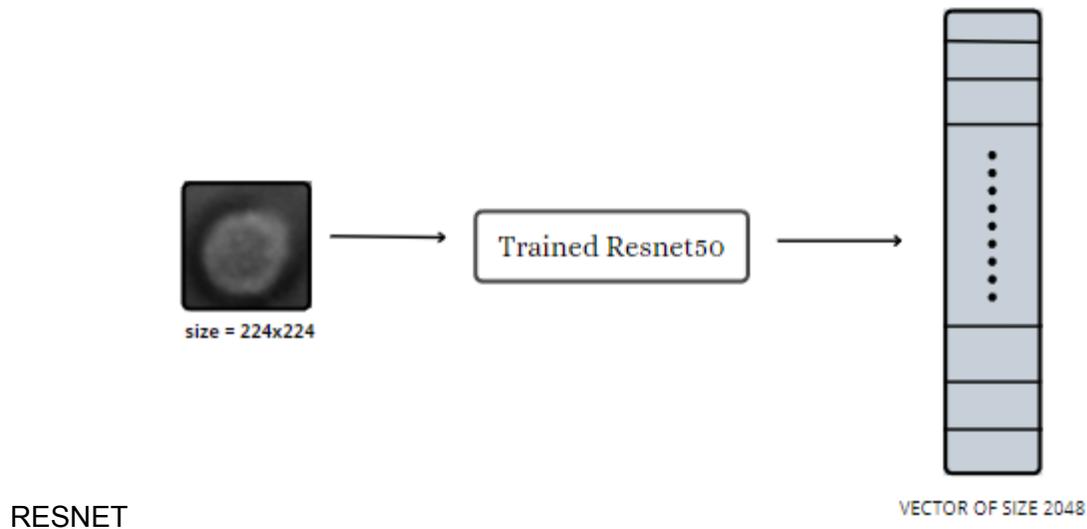

RESNET

**fig(3) Node Feature extraction using Resnet**

**3.3 Extraction of Image embedding using CNN.** Videos are the collection of images and images are the 2-D matrices. Each frame of the video is the collection of one or more than one cell. We have cropped the Target cell from each frame of the video using the tracking coordinates obtained by the TIMING. .We have trained a resnet on the cell images to extract the features from the average pooling layers. Trained Resner is used to convert each cell image into the vector of 2048 and this vector is used as node features for Graph which represent a Frame of the video. Image is resized into the shape of 224x224 before feeding into the resnet architecture as shown in the above fig(3).

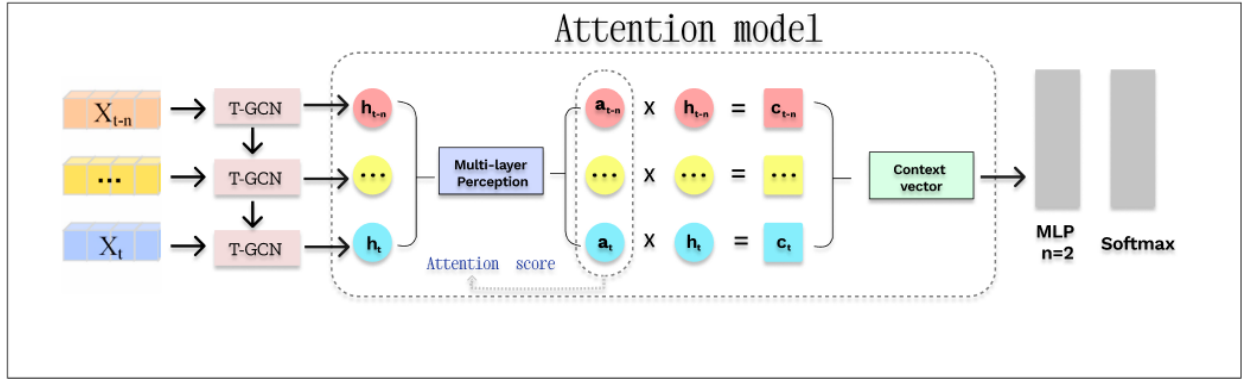

**fig(4) Attention based graph convolution network architecture**

3.4 **Attention based graph neural network:** We have used an Attention based spatio-temporal graph convolutional network[21] as shown in the fig(4) which was developed for traffic flow forecasting to perform the cell death classification. This model has following section as follows :

1) **T-GCN:** A temporal GCN (T-GCN)[22] model is the combination of Graph Convolution network (GCN )[23] and Gated Recurrent Unit (GRU)[24]. Video with n time stamps ( frames ) of cell interaction were inputted into the T-GCN model which gives the n hidden states (h) and these hidden states covered spatiotemporal characteristics:{ht−n,··· , ht−1, ht}. TGCN model calculation is shown in eq. (1). There ht−1 is the output at t-1. GraphConv is the graph convolution. u(t) and r(t) are called the update and reset gates at t.c(t) are called the cell state at the time stamp t. Reset gate, update gate and cell state have learnable parameters.

$$u_t = \sigma(W_u * [GC(A, X_t), h_{t-1}] + b_u) \quad (1)$$
$$r_t = \sigma(W_r * [GC(A, X_t), h_{t-1}] + b_r) \quad (2)$$
$$c_t = tanh(W_c * [GC(A, X_t), (r_t * h_{t-1})] + b_c) \quad (3)$$
$$h_t = u_t * h_{t-1} + (1 - u_t) * c_t \quad (4)$$

The hidden states were calculated for each step and these hidden states were given as input to the attention mechanism to obtain the context vector. Later in the network we used the linear map with two nodes. The output of these two nodes are passed through the softmax to get the probability for each class.

2) **Attention Mechanism:** Attention mechanism was developed as the shortcoming of the encoder decoder architecture. It is used in machine translation to solve the problem of long dependencies. Attention is widely used with images, text data and the time series data. There are many types of attention mechanisms for example, soft attention, hard attention, general attention etc. In this method, the soft attention mechanism is used which means this attention mechanism is differentiable in nature and can be trained using the backpropagation algorithm.

The TGCN model outputs the hidden state for each time step and we use this hidden state output to calculate the alignment score by learning a neural network. We are using a sequence of length 15 and each element of this sequence is a graph and each node of a graph represents the cell image at that particular timestamp t. As discussed above, we have assigned each node with the learned embedding using Resnet. Therefore the input to the TGCN model is the sequence of graphs ( G(1), G(2),G(3),......G(T)). Hidden state is calculated at each time step through TGCN model and then we learn a neural network to calculate the alignment score or the similarity score as follows

$$e_i = w_{(2)}(w_{(1)}H + b_{(1)}) + b_{(2)} \quad (5)$$

The alignment scores are passed to the softmax to calculate the probability distribution.

$$\alpha_i = \frac{\exp(e_i)}{\sum_{k=1}^{n} \exp(e_k)} \quad (6)$$

Context vector is calculated as below

$$C_t = \sum_{i=1}^{n} \alpha_i * h_i \quad (7)$$

We pass this context vector to the MLP with the two neurons and the softmax layer at the end to predict the probability for the cell being dead or alive.

**3.5 Training:** Resnet50 is trained to extract the features for each cell and we have obtained an image embedding vector of 2048 for each image. Attention based temporal graph convolution network is trained on the 122 videos of 15 timestamps each. The class distribution of nodes being assigned labels are shown below in the figure 5 . Cell death is a rare event where the dataset is highly imbalanced. Out of 122 videos, there are 14 videos for which node1 was assigned the death label and node2 was assigned death in 10 videos and node3 was assigned death label around 7 videos. This distribution makes the training dataset quite imbalanced and makes it hard for the model to learn.

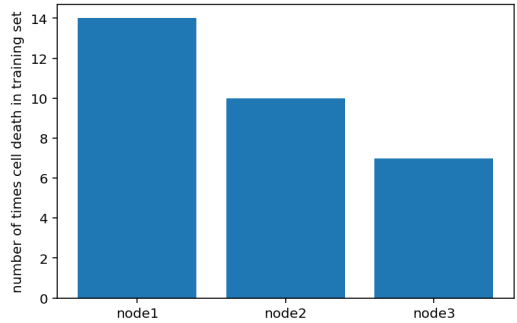

**fig(5) cell death label distribution in training set**

**3.6 Loss function:** Cross entropy loss function is calculated for each node in the graph. Lster , we summed up the losses for each node and used the backpropagation to update the gradient and calculated the weight update based on the total loss calculated.

$$\text{Loss} = \sum_{i=1}^{3} crossentropy\,(y, \hat{y}) \quad (8)$$

**4) Results:** The whole graph network pipeline is tested on the test set having 61 videos with 15 frames. The distribution of the test set is quite imbalanced as shown in fig(). We have calculated the Accuracy, Precision and Recall for each node in the confusion matrix as shown in the below. Node1 is classified well in comparison to the other nodes.

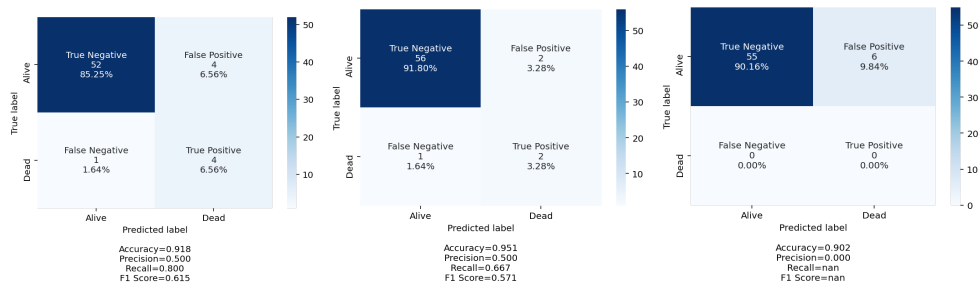

   Node ( 1)          Node ( 2)         Node (3)

In the above Node 3 confusion matrix the precision and recall is nan since node 3 was never assigned a label death. This is due to the dataset in which we do not have three cells in the videos and the 3rd cell having the death label assigned to it.

| Average Accuracy (on two nodes) % | Test Loss | Average Precision (on two nodes) % | Average Recall |
|---|---|---|---|
| 92.36 | 0.0308 | 50.00 | 73.35 |

Table (1): Results for Attention based graph convolution network

Our proposed model is able to achieve good accuracy in comparison to the other state of the art CNN based methods ( Resnet, Alexnet-50) discussed in this paper[25]. But still Recall of the model can be improved to identify the death events.Overall, it shows the graph neural network can be better methods to solve the problem of apoptosis classification due to the cell interaction that happens at each time step.

**5) Future work:** Our method does not account for the effector cell to the target cell interaction right now. We are working on further ideas to encounter the effector cells and trying other graph based networks to get better performance.

**References:**

[1] S. Elmore, "Apoptosis: a review of programmed cell death," Toxicologic pathology, vol. 35, no. 4, pp. 495–516, 2007.

[2] Y. Fuchs and H. Steller, "Programmed cell death in animal development and disease," Cell, vol. 147, no. 4, pp. 742–758, 2011.

[3] A.-L. Nieminen, G. J. Gores, J. M. Bond, R. Imberti, B. Herman, and J. J. Lemasters, "A novel cytotoxicity screening assay using a multiwell fluorescence scanner," Toxicology and applied pharmacology, vol. 115, no. 2, pp. 147–155, 1992.

[4] J. Kuhn, E. Shaffer, J. Mena, B. Breton, J. Parent, B. Rappaz, M. Cham- ¨ bon, Y. Emery, P. Magistretti, C. Depeursinge et al., "Label-free cytotoxicity screening assay by digital holographic microscopy," Assay and drug development technologies, vol. 11, no. 2, pp. 101–107, 2013.

[5] L. L. Chan, S. L. Gosangari, K. L. Watkin, and B. T. Cunningham, "Label-free imaging of cancer cells using photonic crystal biosensors and application to cytotoxicity screening of a natural compound library," Sensors and Actuators B: Chemical, vol. 132, no. 2, pp. 418–425, 2008.

[6] Z. Wang, M.-C. Kim, M. Marquez, and T. Thorsen, "High-density microfluidic arrays for cell cytotoxicity analysis," Lab on a Chip, vol. 7, no. 6, pp. 740–745, 2007.

[7] A. Merouane, N. Rey-Villamizar, Y. Lu, I. Liadi, G. Romain, J. Lu, H. Singh, L. J. Cooper, N. Varadarajan, and B. Roysam, "Automated profiling of individual cell–cell interactions from high-throughput timelapse imaging microscopy in nanowell grids (timing)," Bioinformatics, vol. 31, no. 19, pp. 3189–3197, 2015.

[8] H. Lu, J. Li, M. A. Martinez Paniagua, I. N. Bandey, A. Amritkar, H. Singh, D. Mayerich, N. Varadarajan, B. Roysam, and R. Murphy, "Timing 2.0: High-throughput single-cell profiling of dynamic cellcell interactions by time-lapse imaging microscopy in nanowell grids." Bioinformatics, vol. 1, p. 3, 2018.

[9] C. H. June and M. Sadelain, "Chimeric antigen receptor therapy," New England Journal of Medicine, vol. 379, no. 1, pp. 64–73, 2018.

[10] M. N. Androulla and P. C. Lefkothea, "Car t-cell therapy: A new era in cancer immunotherapy," Current pharmaceutical biotechnology, vol. 19, no. 1, pp. 5–18, 2018.

[11] D. Pischel, J. H. Buchbinder, K. Sundmacher, I. N. Lavrik, and R. J. Flassig, "A guide to automated apoptosis detection: How to make sense of imaging flow cytometry data," PloS one, vol. 13, no. 5, p. e0197208, 2018.

[12] P. Eulenberg, N. Kohler, T. Blasi, A. Filby, A. E. Carpenter, P. Rees, F. J. ¨Theis, and F. A. Wolf, "Reconstructing cell cycle and disease progression using deep learning," Nature communications, vol. 8, no. 1, p. 463, 2017.

[13] T. H. Ward, J. Cummings, E. Dean, A. Greystoke, J.-M. Hou, A. Backen, M. Ranson, and C. Dive, "Biomarkers of apoptosis," British journal of cancer, vol. 99, no. 6, p. 841, 2008.

[14] I. Vermes, C. Haanen, H. Steffens-Nakken, and C. Reutellingsperger, "A novel assay for apoptosis flow cytometric detection of phosphatidylserine expression on early apoptotic cells using fluorescein labelled annexin v," Journal of immunological methods, vol. 184, no. 1, pp. 39–51, 1995.

[15] M. Van Engeland, L. J. Nieland, F. C. Ramaekers, B. Schutte, and C. P. Reutelingsperger, "Annexin v-affinity assay: a review on an apoptosis detection system based on phosphatidylserine exposure," Cytometry: The Journal of the International Society for Analytical Cytology, vol. 31, no. 1, pp. 1–9, 1998.

[16] G. Zhang, V. Gurtu, S. R. Kain, G. Yan et al., "Early detection of apoptosis using a fluorescent conjugate of annexin v," Biotechniques, vol. 23, no. 3, pp. 525–531, 1997.

[17] D. Wlodkowic, J. Skommer, S. Faley, Z. Darzynkiewicz, and J. M. Cooper, "Dynamic analysis of apoptosis using cyanine syto probes: from classical to microfluidic cytometry," Experimental cell research, vol. 315, no. 10, pp. 1706–1714, 2009.
[18] S. Huh, H. Su, T. Kanade et al., "Apoptosis detection for adherent cell populations in time-lapse phase-contrast microscopy images," in International Conference on Medical Image Computing and ComputerAssisted Intervention. Springer, 2012, pp. 331–339.

[19] Manali Singha, Limeng Pu, Abd-El-Monsif Shawky, Konstantin Busch, Hsiao-Chun Wu, J. Ramanujam,  Michal Brylinski: GraphGR: A graph neural network to predict the effect of pharmacotherapy on the cancer cell growth

[20] Kaiming He, Xiangyu Zhang, Shaoqing Ren, Jian Sun:Deep Residual Learning for Image Recognition

[21] Jiawei Zhu, Yujiao Song, Lin Zhao and Haifeng. Li,A3T-GCN: Attention Temporal Graph Convolutional Network for Traffic Forecasting
[22] ] Ling Zhao, Yujiao Song, Chao Zhang, Yu Liu, Pu Wang, Tao Lin, Min Deng, and Haifeng Li. T-gcn: A temporal graph convolutional network for traffic prediction. IEEE Transactions on Intelligent Transportation Systems, 2019.

[23] Thomas N. Kipf and Max Welling. Semi-supervised classification with graph convolutional networks, 2016

[24] Kyunghyun Cho, Bart Van Merrienboer, Dzmitry Bahdanau, and Yoshua Bengio. On the properties of neural machine translation: Encoder-decoder approaches. Computer Science, 2014.

[25] Aryan Mobiny, Hengyang Lu, Hien V. Nguyen,, Badrinath Roysam, Navin Varadarajan:Automated Classification of Apoptosis in Phase Contrast Microscopy Using Capsule Network,2019